\documentstyle[prb,twocolumn,aps]{revtex}

\begin{document}
\tolerance 50000

\draft

\title{Optical Conductivity of a t--J Ladder}
\author{C.A.Hayward$^{1}$, D.Poilblanc$^{1,2}$ \&
 D.J.Scalapino$^{2}$
}
\address{
$^{1}$Lab. de Physique Quantique, Universit\'e Paul Sabatier,
31062 Toulouse, France \\
$^{2}$Department of Physics, University of California Santa Barbara, CA
93106-9530 \\
}

\twocolumn[
\date{December 95}
\maketitle
\widetext

\vspace*{-1.0truecm}

\begin{abstract}
\begin{center}
\parbox{14cm}{
The optical conductivity $\sigma(\omega )$ of a doped two-leg t--J ladder
is calculated for an electric field polarized parallel to the legs of the
ladder.	The conductivity has a Drude weight proportional to the hole doping
and an apparent threshold for absorption (a pseudo gap) which may be associated
with the energy to break a pair. This pseudogap
in $\sigma (\omega )$ is present
even though the pairs have a modified $d_{x^2-y^2}$-like
wave function because the geometry of the ladder leads to quasi-particle states
which probe the gap along an antinode.
}
\end{center}
\end{abstract}

\pacs{
\hspace{1.9cm}
}
]
\narrowtext

Theoretical predictions \cite{heisladder} of the existence of a spin gap in a
two-leg
$S=1/2$ antiferromagnetic ladder have now been observed in spin susceptibility
measurements of vanadyl pyrophosphate \cite{compounds1} $(VO)_2P_2O_7$
and the cuprate material \cite{compounds2} $SrCu_2O_3$. These materials are
insulating so that the charge excitations are gapped as well as the spin
excitations. Recently, Hiroi and Takano \cite{compounds3} have  succeeded in
doping a
two-leg ladder compound $La_{1-x}Sr_xCuO_{2.5}$, and observed an insulator to
metal transition \cite{note1}. Numerical \cite{PSH,HPNSH,HP} and analytic
\cite{TTR,NS}
calculations have found that hole-doped two-leg t--J ladders
have a gap to spin excitations
but no gap to charge excitations. These calculations show that the doped holes
form modified '$d_{x^2-y^2}$-like' pairs and that the pair-field correlations
have
a power-law decay.
In addition the pairs are expected to exhibit power-law '$4k_f$' charge-density
correlations.

The absence of a charge gap implies the existence of a finite Drude
weight which is expected to be proportional to the hole doping at small doping
levels. In addition, the possibility of breaking up the pairs suggests that in
spite of the vanishing of the charge gap reflected in the collective motion
of the pairs, there should be a pseudogap in the frequency dependence of the
charge excitations associated with breaking the pairs (we note that
Giamarchi {\it et al} \cite{GM}
have predicted $\omega^3$ behaviour for the low frequency
conductivity of a Luttinger Liquid, although
direct calculations of this behaviour are as yet
unobserved).
Here we seek a further understanding of the charge response of the doped
two-leg t--J
ladders by examining the frequency-dependent conductivity
$\sigma (\omega )$ for an E-field polarized along the legs of the ladder.
Lanczos
calculations for  doped $2\times 5$ and $2\times 10$ t--J ladder have been
carried
out and the behaviour of $\sigma (\omega )$ has been investigated.

The t--J hamiltonian on a $2\times L$ ladder is defined as
\begin{eqnarray}
   {\cal H}=
   J^\prime \; \sum_{j}
   ({\bf S}_{j,1} \cdot {\bf S}_{j,2}
   - \textstyle{1\over4} n_{j,1} n_{j,2} ) \cr
   +J\;\sum_{\beta,j}
   ({\bf S}_{j,\beta} \cdot {\bf S}_{j+1,\beta}
   - \textstyle{1\over4} n_{j,\beta} n_{j+1,\beta} ) \cr
      - {t} \sum_{j,\beta ,s}
   P_G({c}^{\dagger}_{j,\beta;s}{c}_{j+1,\beta;s} + {\rm H.c.})P_G  \cr
      - {t^\prime} \sum_{j,s}
   P_G({c}^{\dagger}_{j,1;s}{c}_{j,2;s} + {\rm H.c.})P_G
\label{hamiltonian}
\end{eqnarray}
\noindent
where $\beta$ (=1,2)
labels the two legs of the ladder (oriented along the
$x$-axis) and $j$ is a rung index ($j$=1,...,$L$).
${\bf S}_{j,\beta}$ and $c^\dagger_{j,\beta;s}$ are electron spin and creation
operators and the Gutzwiller projector $P_G$ excludes configurations
with doubly occupied sites. In the materials of interest, the exchange
coupling $J$ along the legs is nearly the same as the exchange
coupling $J^\prime$ across a rung, and similarly the hopping $t$ along
the legs is close to the rung hopping strength $t^\prime$;
therefore, in the following calculations
we will work with the isotropic system,  $J=J^\prime$ and $t=t^\prime$.

For an electric field polarized along the $x-$axis (parallel to the legs),
the optical conductivity can be expressed as
\begin{eqnarray}
\sigma (\omega )=
2\pi D\delta (\omega )&
\cr
+{\pi\over N} \sum_{n(\neq 0)} { |{
\langle \phi_n^M| j_x | \phi_0^M\rangle
}|^2 \over {E_n^M-E_0^M}}& \delta (\omega-E_n^M+E_0^M)
\label{conductivity}
\end{eqnarray}
\noindent
where the final term is usually defined as the regular part of the conductivity
$(\sigma^{reg} (\omega ))$. $D$ is the charge stiffness and
$j_x$ is the paramagnetic current operator
\begin{eqnarray}
j_x=-it\sum_{j,\beta;s} (c^\dagger_{j+x,\beta;s}
c_{j,\beta;s}-c^\dagger_{j,\beta;s}c_{j+x,\beta;s})
\label{current}
\end{eqnarray}
\noindent
where we choose the charge $e$ to equal unity.
$N=2L$ is the number of sites and M is the number of holes in the ladder
(assumed to be even throughout). The states $| \phi_n^M\rangle$
in Eq. \ref{conductivity} are the energy eigenstates of the
hamiltonian ${\cal H}$  with corresponding
energies $E_n^M$; $| \phi_0^M\rangle$ is the ground state with energy $E_0^M$.

The Drude weight $\sigma_0=2\pi D$
may be evaluated in two ways.
Firstly, as explained in a previous publication \cite{HP},  the charge
stiffness $D$ can be
calculated by considering the curvature of the ground state energy level
($E_0^M$)
as a function of the flux  $\Phi$ (in units of the flux quantum
$\Phi_0=hc/e$) threaded through the two chain system
\begin{eqnarray}
D={L^2\over 8\pi^2}{\partial^2(E_0^M/N)\over \partial \Phi^2}
\label{drudecurv}
\end{eqnarray}
\noindent
Secondly, $\sigma_0$ may be calculated using the sum rule
\cite{MD,D}
\begin{eqnarray}
{N\over 2\pi}\int_{-\infty}^\infty \sigma (\omega ) d\omega = {1\over 2}
|\langle \phi_0^M | T_{xx} | \phi_0^M\rangle|
\label{sumrule}
\end{eqnarray}
\noindent
where the Kinetic Energy operator $T_{xx}$ along the chain direction
consists of the term proportional to t in
the hamiltonian (\ref{hamiltonian}). Substituting Eq. \ref{conductivity}
into Eq. \ref{sumrule}, we have
\begin{eqnarray}
{N\sigma_0\over 2\pi}
+\sum_{n(\neq 0)}{| \langle \phi_n^M|j_x|\phi_0^M\rangle |^2\over E_n^M-E_0^M}
={1\over 2}|\langle \phi_0^M | T_{xx} | \phi_0^M \rangle |
\label{sumrule2}
\end{eqnarray}
\noindent
where the second term is equal to $(N/\pi )\int^\infty_0 \sigma^{reg}
(\omega ) d\omega $.

In Fig. \ref{f1} we plot  ${N}\times \sigma^{reg}(\omega )$
for the
ladder system with various values of the
ratio $J/t$: The results were  obtained using the Lanczos approach
and data
from both the $2\times 5$ and $2\times 10$ system are shown.
In order to obtain the absolute ground state
of the system, we have chosen boundary conditions to form a closed
shell in the non-interacting Fermi sea (obtained by turning off the
interactions $J$ and $J^\prime$): specifically this
corresponds to anti-periodic boundary conditions for
$n< 0.5$ and periodic boundary conditions for $n>0.5$.
We have also considered the parity of the states under a reflection in the
symmetry axis of the ladder along the direction of the chains (even ($R_x=1$)
or odd ($R_x=-1$)).

In Fig. \ref{f2}a we plot the Drude weight $\sigma_0$ as a function of
electron density for various values of the ratio $J/t$; we  note that both
of the methods above were used  (Eqs. \ref{drudecurv} and \ref{sumrule2})
and the results were in excellent agreement.
In Fig. \ref{f2}b we plot the integrated finite frequency conductivity
$2\int^\infty_{0}d\omega\sigma^{reg} (\omega )$ as a function of electron
density for various values of the ratio $J/t$; this quantity can be directly
related
to $\sigma_0$ and to $\pi$ times the kinetic energy per site through
the sum rule (Eqs. \ref{sumrule} and \ref{sumrule2}).

There are several important features of Figs. \ref{f1} and \ref{f2} we
should mention.
We have checked that as the system size is increased,
the ratio of weight at finite frequency to the total conductivity remains
effectively
constant; the difference in the number of delta peaks as the system size  is
changed is
merely a finite size effect. We have also checked that the moments of the
finite frequency distribution
are effectively unchanged with increasing system size.
A `pseudogap' appears in the optical conductivity, below which there
appears to be no weight; we discuss this feature in some detail below;
note that as the ratio
$J/t$ is increased, the `pseudogap' increases.
Secondly, as can be seen by comparing Figs. \ref{f2}a and b,
the weight of the conductivity at finite frequency
is extremely small as compared to that at zero frequency (the Drude weight).
This behaviour is reminiscent of the one-dimensional t--J model and
in contrast to the larger finite frequency weight found in the two-dimensional
system \cite{D}. The Drude weight is proportional to the hole-doping at low
doping levels, and at low densities is effectively independent
of $J$ as we would expect for a spin-charge separated state.
Note however, that the small finite frequency absorption has a larger magnitude
together with  a larger and {\it opposite} J dependence for $n>0.5$.
These differences might be related to the nature of the ground state,
a Tomonaga Luttinger liquid at low electron density and a Luther-Emery like
liquid
with a spin gap at small hole density.

In picturing the intermediate states which enter Eq. \ref{conductivity}
for the conductivity, it is convenient to consider the limit in which the
rung exchange $J^\prime$ is large compared to both $J$ and the hopping
$t$(=$t^\prime$). In this case, when two holes are doped into the
antiferromagnetic
ladder they will go onto the same rung in order to minimize the number
of broken singlet rung bonds.  The binding energy of this state is
approximately
$J^\prime-2t-2t^\prime$, reflecting
$J^\prime$ the energy required to break another rung singlet minus
the kinetic energy gain for two separate holes. In the limit of large
$J^\prime$,
the doped system can therefore be thought of in terms of hole pairs moving on a
lattice of rung singlets with an effective pair transfer matrix element of
order $-2t^2/J^\prime$. The fact that only one pair of holes can
occupy a given rung leads to a low-energy description in terms of
hard-core charge 2e bosons.

The coherent propagation of pairs leads to a finite Drude weight. In addition,
when $\omega$ exceeds an energy of the order of the pair binding energy $E_B$,
it is possible for the system to absorb a photon and excite a state containing
two
quasi-particles (each with charge $e$ and spin $1/2$).
The two quasi-particle state which enters Eq. \ref{conductivity}
for $\sigma (\omega )$ has S=0 and a center of mass momentum equal
to zero. There is in fact a continuum of two quasi-particle scattering
states  above the threshold energy for breaking a pair in the
infinite ladder. As discussed by Tsunetsugu {\it et al} \cite{TTR}, a
similar two quasi-particle state with $S=1$ determines the spin gap in the
doped
two-leg ladder.
An estimate of the pair binding energy can be obtained by
considering the ground state energy
of a system with an even number of holes $M$ and the corresponding energy if a
hole
with a momemtum k along the chains is either added or removed.
Such excitations are infact the quantities we would expect to measure
in the spectral functions,
$A({\bf k},\omega )=\sum_n
|\langle \phi_n^{M-1} | c^\dagger_{k,s}|\phi_0^M\rangle |^2 \delta (\omega-
E_n^{M-1}+E_0^M)$ and
$A({\bf k},\omega )=\sum_n
|\langle \phi_n^{M+1} | c_{k,s}|\phi_0^M\rangle |^2 \delta (\omega+
E_n^{M+1}-E_0^M)$ where the first expression relates to the addition of an
electron
and the second relates to the addition of a hole.
Indeed, sharp features in the spectral function should appear at energies
$e^\pm_{k,R_x}=\pm(E_0^{M}-E_0^{M\pm 1}(k,R_x))$
where we can obtain two different values depending whether we look at bonding
or antibonding excitations  ($R_x=\pm 1$) \cite{note2}.

In order to consider in more detail the quasi-particle excitations
we define the (average) chemical potential as
$\mu$=$1/2(E_0^{M+1}-E_0^{M-1}) $
where $E_0^P$ relates now to the absolute ground state energy for P holes
irrespective of the quantum numbers k and $R_x$.
In Fig. \ref{f3} we show the dispersions $e^\pm_{k,R_x}-\mu$ for the $2\times
10$ ladder
at $n=0.8$ and $J/t=1.0$ with positive energies corresponding to the case of
adding an
electron to the 4 hole ground state and negative energies corresponding to
adding
a hole to the 4 hole case.
A finite gap separating the electron and hole quasi particle spectra can be
clearly seen on the figure
both for the bonding or antibonding sectors.
Indeed, these gaps are related to the binding energies of a pair,
\begin{eqnarray}
E_B(R_x)=E_0^{M-1}(R_x)+E_0^{M+1}(R_x)-2 E_0^{M},
\label{binding}
\end{eqnarray}
\noindent
i.e. the energy required to create an electron-hole excitation with $R_x=\pm
1$.
These two processes correspond in Fig. \ref{f3} to transitions involving
the smallest energy excitations as indicated by the arrows.

Since these transitions conserve momentum (not quite exactly for $R_x=-1$) they
can be considered
as 'vertical' optical transitions and should be seen in the optical absorption
(\ref{conductivity}).
In  Fig. \ref{f4} we compare the binding energies, the pseudogap
obtained from the optical conductivity data and the two spin gaps
(corresponding to bonding and anti-bonding spin excitations) at $n=0.8$
(i.e $N_h=4$) for various
ratios of the parameter $J/t$ in the
$2\times 10$ ladder.
As  expected the threshold for absorption is of the order of the
pair binding energy and also provides a measure of the spin gap in the
$R_x=1$ spin excitation spectrum.
This adds credit to our picture of the pseudogap being a threshold for
unbinding pairs.
It is interesting to note that other lower energy gapped
spin excitations exist (with $R_x=-1$) which probably do not involve pair
breaking.

In conclusion, we find that $\sigma (\omega )$ for the two leg ladder has a
Drude
contribution consistent with a coherent pair motion and an absorption onset
which
we associate with the creation of two quasi-particles. It is interesting to
note
that for the ladder there is a pseudo-gap in $\sigma (\omega )$ even though the
pairs have a $d_{x^2-y^2}$-like wavefunction. Basically, one is confined by
the geometry to probe the gap along the antinode so that $\sigma (\omega )$
has a finite threshold.

\bigskip
We gratefully acknowledge many useful discussions with
M. Luchini and F. Mila.
{\it Laboratoire de Physique Quantique, Toulouse} is
{\it Unit\'e de Recherche Associ\'e au CNRS No 505}.
CAH and
DP  acknowledge support from the EEC Human Capital and Mobility
program under Grants ERBCHBICT941392 and  CHRX-CT93-0332; DJS  acknowledges
support from the Department of Energy under grant DE FG03 85ER45197 and the
Program
on Correlated Electrons at the Center for Materials Science at Los Alamos
National Laboratory.
We also thank IDRIS (Orsay)
for allocation of CPU time on the C94 and C98 CRAY supercomputers.

%
%
\begin{figure}
\caption{
$N\times$ $\sigma^{reg}(\omega )$ for  $2\times 5$  and
$2\times 10$ ladders at electron density $n=0.8$. The different plots
correspond
to a) J/t=0.5 b)J/t=1.0 c)J/t=1.5
\label{f1}
}
\end{figure}

%
%
\begin{figure}
\caption{
a) The Drude weight $\sigma_0$, and
b) The sum of the finite frequncy conductivity
$2\int^\infty_{0^+}\sigma (\omega )$,
versus electron density for various values of $J/t$.
These two quantities are related by the sum rule (see text).
\label{f2}
}
\end{figure}

%
%
\begin{figure}
\caption{
Energy dispersion for the $2\times 10$ ladder system ($J/t=1.0$) with 3 holes
(upper curves)
and 5 holes (lower curves); the results for both $R_x=1$ and $R_x=-1$ are
shown. Momentum is in units of $\pi$. The arrows represent the energy required
to
create an electron-hole excitation with either $R_x=\pm 1$ (see text).
\label{f3}
}
\end{figure}

%
%
\begin{figure}
\caption{
A comparison of the the binding energy, the pseudogap (obtained
from the optical conductivity data) and the spin gaps. These quantities
are defined in the text.
\label{f4}
}
\end{figure}


\begin{references}

\bibitem{heisladder} T.Barnes {\it et al}, Phys.Rev.B {\bf 47}, 3196 (1993);
     S.Gopalan, T,M.Rice and M.Sigrist, Phys.Rev.B {\bf 49}, 8901 (1994).
\bibitem{compounds1} D.C.Johnston {\it et al}, Phys.Rev.B {\bf 35}, 219 (1987)
\bibitem{compounds2} Z.Hiroi {\it et al}, Physica C {\bf 185-189}, 523 (1991);
M.Takano {\it et al}, JJAP series {\bf 7}, 3 (1992)
\bibitem{compounds3} Z.Hiroi and M.Takano, {\it Nature} {\bf 377}, 41 (1995).
\bibitem{note1} However, band structure calculations indicate that the
ladders in this material may in fact be rather strongly coupled.
\bibitem{PSH} D.Poilblanc, D.J.Scalapino and W.Hanke, Phys.Rev.B. {\bf 52} 6796
(1995).
\bibitem{HPNSH} C.A.Hayward {\it et al}, Phys.Rev.Lett. {\bf 75} 926 (1995)
\bibitem{HP} C.A.Hayward and D.Poilblanc, submitted to Phys.Rev.B (Sept 1995).
\bibitem{TTR} H.Tsunetsugu, M.Troyer and T.M.Rice, Phys.Rev.B {\bf 49} 16078
(1994);
Phys.Rev.B {\bf 51} 16456 (1995).
\bibitem{NS} N.Nagaosa, preprint; H.J.Schulz, Phys.Rev.B, to be published
(Feb.1996)
\bibitem{GM} T.Giamarchi and A.J.Millis, Phys.Rev.B {\bf 46} 9325 (1992)
\bibitem{MD} P.F.Maldagne, Phys.Rev.B. {\bf 16} 2437 (1977); see also
D.Poilblanc, Phys.Rev.B {\bf 44} 9562 (1991).
\bibitem{D}
D.Poilblanc and E.Dagotto, Phys.Rev.B {\bf 44} 466 (1991);
P.Beran, D.Poilblanc and R.B.Laughlin, submitted to Nuclear Phys. B.
\bibitem{note2} A calculation of the relative weights of the quasi-particle
excitations
is left as a future study.
\end{references}
\end{document}